 \documentclass[final,1p,times]{elsarticle}

 \usepackage{graphics}
 \usepackage{graphicx}
 \usepackage{epsfig}
\usepackage{amssymb}
\usepackage{amsthm}
\def\astrobj#1{SDSS J092009.54+004244.9}
\def\astrobj#1{SDSS J132723.39+652854.3}
\def\astrobj#1{SDSS J122740.83+513925.0}
\def\astrobj#1{SDSS J160745.02+362320.7}
\def\astrobj#1{SDSS J145758.21+514807.9}
\def\astrobj#1{V849 Ophiuchi}

\journal{New Astronomy}
\begin{document}
\begin{frontmatter}

  \title{Observations of Faint Eclipsing Cataclysmic Variables\tnoteref{t1}}
  \tnotetext[t1]{Based on photometric observations performed at T\"{U}B\.{I}TAK National Observatory (Turkey)}
  \author[rvt]{D.~Zengin \c{C}amurdan\corref{cor1}}
  \ead{dicle.zengincamurdan@ege.edu.tr}
      \author[rvt]{C. \.{I}bano\v{g}lu}
      \author[rvt]{C.M.~\c{C}amurdan}
       
          \cortext[cor1]{Corresponding author}
\address[rvt]{Ege University, Science Faculty, Department of Astronomy and Space Sciences, 35100 Bornova, \.{I}zmir, Turkey
}
   \begin{abstract}
We present time-resolved photometry of six faint ($V>17{mag}$) cataclysmic variables (CVs); one of them is V849 Oph and the others are identified from the Sloan Digital Sky Survey (SDSS J0920+0042, SDSS J1327+6528, SDSS J1227+5139, SDSS J1607.02+3623, SDSS J1457+5148). The optical CCD photometric observations of these objects were performed at the T\"{U}B\.{I}TAK National Observatory (Turkey) between February 2006 and March 2009. We aimed to detect short time scale orbital variability arisen from hot-spot modulation, flickering structures which occur from rapid fluctuations of material transferring from red star to white dwarf and orbital period changes for selected short-period ($P<4{h}$) CVs at quiescence. Results obtained from eclipse timings and light curves morphology related to white dwarf stars, accretion disks and hot-spots are discussed for each system. Analysis of the short time coverage of data, obtained for SDSS J1227+5139 indicates a cyclical period change arisen from magnetic activity on the secondary star. Photometric period of SDSS J1607+3623 is derived firstly in this study, while for the other five systems light elements are corrected using the previous and new photometric observations. The nature of SDSS J1457+5148 is not precisely revealed that its light curve shows any periodicity that could be related to the orbital period.
\end{abstract}

\begin{keyword}
stars:cataclysmic variables \sep stars:binaries: eclipsing-stars
\PACS 97.80.Gm \sep 97.80.Hn 
\end{keyword}

\end{frontmatter}


\section{Introduction}
CVs are binary stars containing a white dwarf which accretes matter from a red dwarf companion. Unless the white dwarf is strongly magnetic the mass transferred from the secondary is usually accumulated in a disk.  Completed or ongoing large-scale surveys, such as Sloan Digital Sky Survey (here after SDSS; Szkody et al. 2002), Hamburg Quasar Survey (HQS; G\"{a}nsicke et al. 2002), Faint Sky Variability Survey (FSVS; Groot et al. 2003) provide discovery of a substantial number of sub-types of CVs. In particular, SDSS has revealed a large population of CVs ($\sim$220 systems) with short orbital periods and low mass transfer rates (Szkody et al. 2002, 2003, 2004, 2005, 2006, 2007;  Schmidth et al. 2007) of which 180 are new discoveries. Light curves and time-resolved spectra are being acquired for these objects to identify their orbital periods and to determine their sub-types.

The aims of the present study are to obtain photometry of short-period, faint eclipsing CVs at quiescence and  to search orbital variability due to hot-spot modulation, flickering structures, occurring on short time scales and periodicities related to eclipsing other components (disc, white dwarf). Most of the CVs are too faint for spectroscopic studies by using moderate-sized telescopes, therefore, the mostly used way to determine their orbital periods is timings of photometric eclipses. Time-resolved photometry using  a CCD  allows us to carry out accurate photometric observations even with small telescopes for relatively faint stars ($V>17{mag}$) . Eclipsing systems offer the best opportunities for determining the fundamental properties of cataclysmic variables. Moreover, there are a wide range of photometric studies of short-period CVs to deduce varying structural features either at quiescence or outburst (e.g. Woudt\&Warner 2001, 2002, 2003, Woudt et al. 2004,2005; Howell\&Szkody 1988, Howell et al. 1989, 1990, 1991; Dillon et al. 2008). The targets taken in this study either have little observations or no previous detailed studies.  In this paper, we present time-resolved CCD photometry of the short-period CVs:  \astrobj{SDSS J0920+0042}, 
 SDSS J1327+6528, SDSS J1227+5139, SDSS J1607+3623,  SDSS J1457+5148, V849 Ophiuchi (hereafter SDSS J0920, J1327, J1227, J1607, J1457 and V849 Oph, respectively). The wide-band light curves are shown and the preliminary results are given.
  
\begin{table}
\footnotesize
\begin{minipage}{8cm}
\caption{Log of observations.\label{tbl-1}}
\begin{tabular}{@{}lcccccccccc@{}}
\hline
Object&$P_{orb}$&Type&\textit{g}\footnote[1]{The g magnitudes from the SDSS data (in the case of V849 Oph, Johnson $V$ magnitude)} &Date of Obs.&HJD of first obs.&Filter&Exposures&N\footnote[2]{Number of observations} &T\footnote[3]{Total observation interval in hour} &sigma\footnote[4]{The standard deviation in mag}\\
&(h) &&(mag) &&(+2450000.0)& &(s)& & &(mag)\\
\hline
SDSSJ092009.54+004244.9&3.6& &17.45&2006 Feb 27&3794.30699&$R,V$&200,150&19&2.73&0.02,0.02\\
SDSSJ132723.39+652854.3&3.28&SW Sex&17.77&2007 May 16&4237.29642&Clear&60&151&3.32&0.02\\
& & & & 2008 Mar 03&4529.21983&$V$&100,200&135&5.09&0.04\\
SDSSJ122740.83+513925.0&1.50& &19.10&2008 Mar 04&4530.23094&$V$&200&63 &4.2 &0.09\\
& & & & 2009 Feb 26&4889.32249&$V$&200&136&8.05&0.03\\
SDSSJ160745.02+362320.7&3.49& &18.08&2008 Mar 04&4530.41773&$V$&100&112&3.56&0.05\\
SDSSJ145758.21+514807.9& & &19.54&2008 Mar 03&4529.44323&$V$&200&88&4.76&0.06\\
V849 Oph&4.15&CN\footnote[5]{Classical Nova} &17.9&2006 Aug 14&3962.31371&$B,V$&150,120&25&3.24&0.05,0.03\\
& & & &2006 Aug 15&3963.25825&$V$&120&103&4.61&0.03\\
& & & &2006 Aug 16&3964.26426&$B,V$&150,120&41&4.24&0.03,0.03\\
\hline
\bigskip
\end{tabular}
\vspace{-0.9cm}
\end{minipage}
\end{table}

  \section{Observations and Data Analysis}
Table 1 gives journal of observations for six CVs observed by us. The  photometric observations of these objects were performed by the Russian-Turkish 1.5-m telescope RTT150 at the T\"{U}B\.{I}TAK National Observatory (Turkey) using ANDOR CCD (2048$\times$2048 pixels at 0.24 arc s pixel$ ^{-1}$ resolution). We used standard routines for bias subtraction and flat-fielding of the CCD frames using the APPHOT package in IRAF. Differential magnitudes were measured relative to one or more nearby comparison stars on the CCD frame. Since our major goal is to obtain differential photometry for program stars, we did not observe photometric standard stars during our runs. All observations were acquired through Johnson $B,V, R$ bands and clear, i.e., without filter.

\section{Results}
In this section we present the result of photometric observation for each system.

\subsection{SDSS J0920}
This object is one of the CVs discovered by Szkody et al. (2003) from the SDSS spectra obtained in 2001. The eclipsing nature has also been revealed by the same authors. The follow-up photometric observations showed, about two mag, deeply eclipsing system. Most remarkable features in the quiescence spectra (covering the range of 4000-8000 \AA) obtained by Szkody et al. were strong narrow emission lines which do not show the usual deep doubling. There was no trace of secondary star in the red part of the spectrum. The first photometric study of SDSS J0920 was carried out using 1 m telescope, a CCD camera and the Johnson $V$  filter at the US Naval Observatory Flagstaff Station (NOFS). This observation revealed the eclipses of the system with an orbital period of 3.6 h at a quiescence. 

\begin{figure}
\centering
\rotatebox{270}{\includegraphics[scale=0.5]{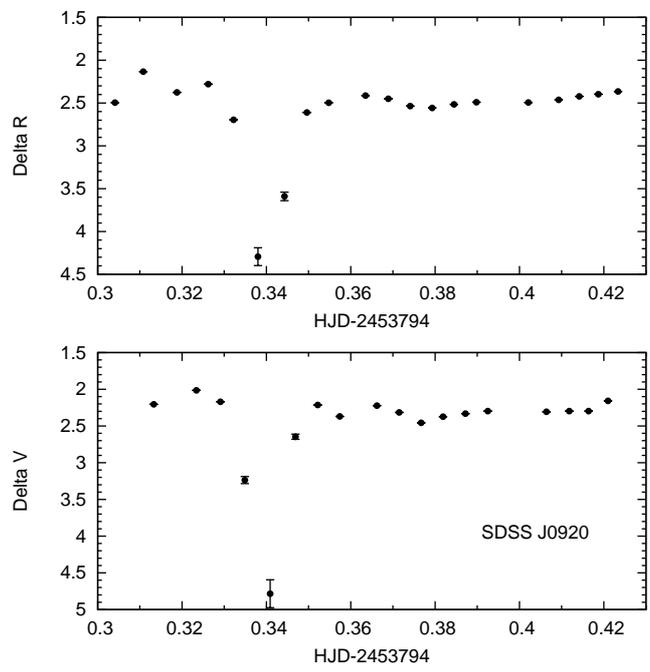}}
\caption{The observed light curves of SDSS J0920 obtained in $R$ filter (top) and in $V$ (bottom) on 2006 Feb 27 at quiescence.  Very few observations could be obtained within the eclipse due to long exposure times used. Each point represents a differential magnitude  relative to a comparison star included in the same CCD frame. An estimate of the photometric accuracy in the differential data is also listed in Table \ref{tbl-1}.}
\label{fig1}
\end{figure}

In 2006 February we started for the  photometric observations of the system in Johnson $V$ and $R$-bandpasses at a quiescence state. A deep eclipse of $\sim$2.5 mag in $V$ filter and $\sim$2.0 mag in $R$ filter are apparent in the light curve of SDSS J0920 (see Fig. \ref{fig1}). An estimate of the photometric accuracy in the differential observation is also listed in Table 1.The depth of the eclipse may be estimated as quite large due to the few points near the bottom of the light curves when the star was too faint to measure. The light curves show no evidence of a hump-like structure that would signify a strong hot spot. The eclipse in the $R$-bandpass appears to be shallower resulting from a cooler donor while large amount of the light in the $V$-bandpass comes from a smaller hot spot. We should mention that the system could not be observed simultaneously in two colours.  Details of the light curves, i.e. light fluctuations in short time intervals,  could not be revealed due to the long exposure times we used  (200 s and 150 s). In other words, the light variations originated from hot spot and mass transfer and as well as eclipse timings in the case of SDSS J0920 could not easily be resolved. 

\subsection{SDSS J1327}
SDSS J1327 was first identified spectroscopically in the SDSS database by Szkody et al. (2003) as a CV and it was classified as an SW Sex star by Wolfe et al. (2003). The system shows a typical strong single-peaked He II $\lambda$4686 emission line. So Wolfe et al. confirmed the earlier classification of this system as a SW Sex star using the results obtained spectroscopic, spectropolarimetric observations and Doppler tomography. Furthermore, the light curve of SDSS J1327 displays deep eclipses with an orbital period of 3.28 h but shows no evidence of pre-eclipse hump originated hot spot. 

We observed SDSS 1327 on 2007 May 16 and and 2008 March 3 with time intervals of 3.32 and 5.09 h, respectively. During the first run, we made observations with clear filter using 60-s integration time. We obtained an eclipse with a depth of $\sim$1.8 mag.  Few  observations were obtained in the bottom of the eclipse due to the long exposure times.

\begin{figure}
\centering
\rotatebox{270}{\includegraphics[scale=0.40]{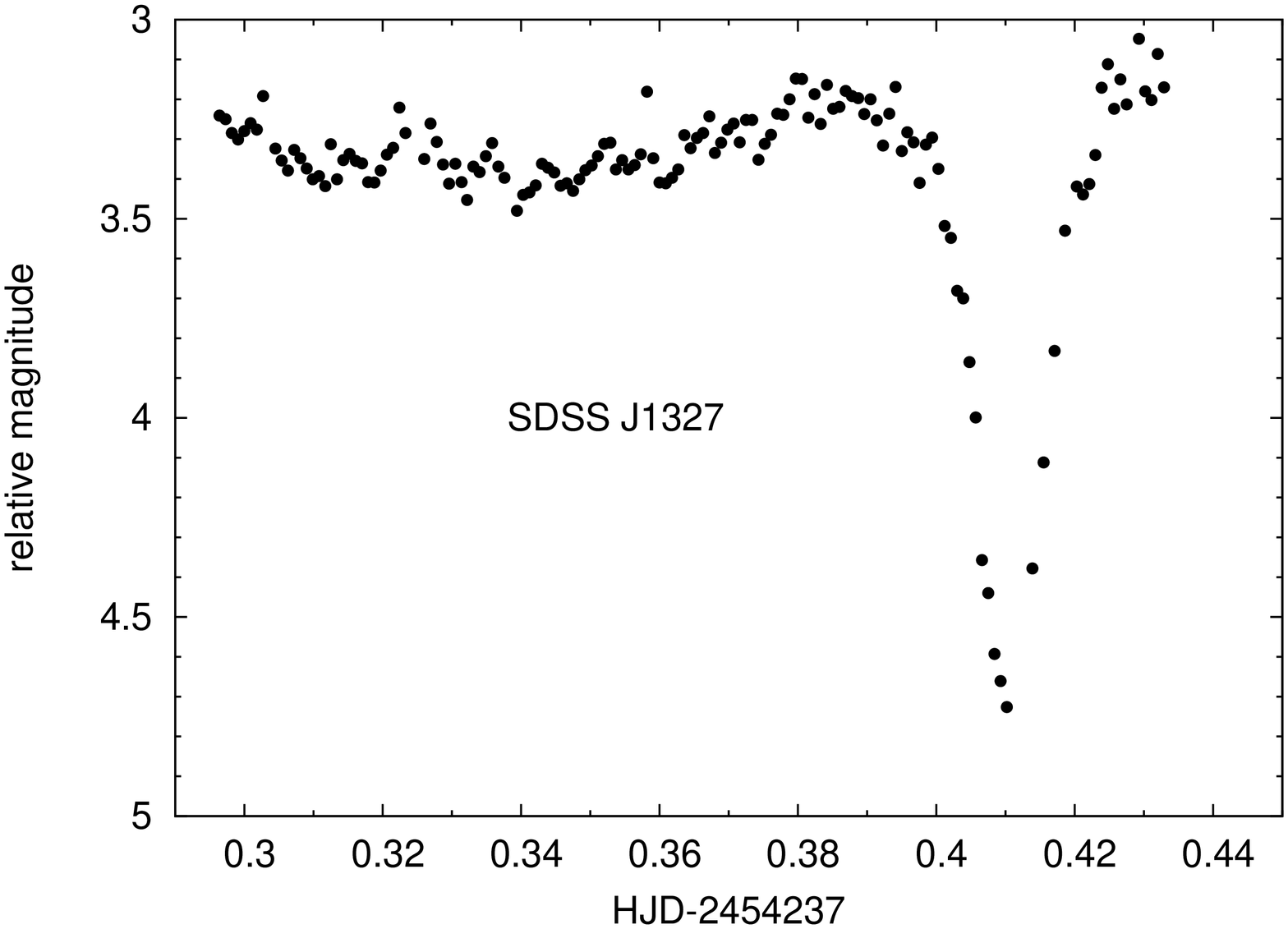}}
\hfill    
\rotatebox{270}{\includegraphics[scale=0.45]{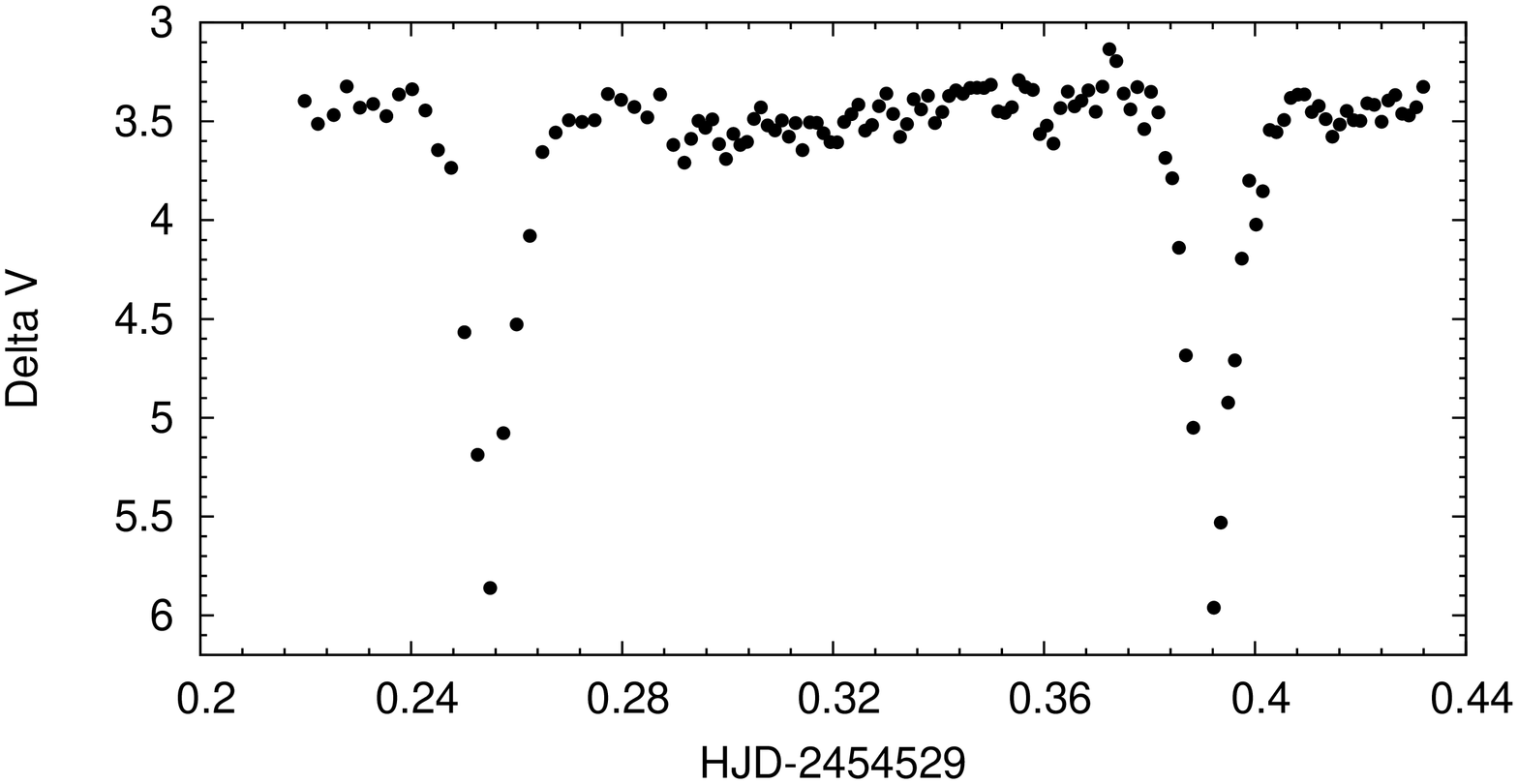}} 
\caption{The light curve of SDSS J1327 obtained without a filter on 2007 May 16 (top) and with \textit{V} filter on 2008 March 3 (bottom). Note the symmetry of the eclipse. The flickering features obviously appear with a variation of 0.15 mag in the first run and 0.25 mag in the second run with a time scale of about $\sim$ 600 s at out-of-eclipse. See Fig. \ref{fig1} caption for a complete description.}
\label{fig2}
\end{figure}

During the second run, we used the $V$ filter of the $UBV$ system and obtained an eclipse with a depth of $\sim$2.5 mag. Fig. \ref{fig2} shows the differential observations with unfiltered and in the $V$-bandpass as a function of the time. The first light curves show an eclipse which is fairly symmetrical in shape and a poor evidence about the existence of a hot spot and disk. There is no sign of a dominant hot spot such as a hump preceding the eclipse in the light curve. As seen in Fig. \ref{fig2}, we noticed that the eclipse depth is highly variable. While the depth was measured by Wolfe et al. as 1.65 mag we estimate it as $\sim$1.8 mag in 2007 and $\sim$2.5 mag in a year later.  Although there is a phase-dependent variation with an orbital  period of 3.28 h, flickering, which could be arisen from turbulent inner disc or the bright spot, can be seen with a short time scale of about $\sim$ 600 s at out-of-eclipse. In Fig. \ref{fig2} the flickering features obviously appear with brightness variation of 0.15 mag on 2007 May 16 and 0.25 mag on 2008 March 3. A linear least-squares fit to the three eclipse timings listed in Table \ref{tbl-2} yielded the following ephemeris:

\begin{equation}
\textnormal{$Min I$(HJD)}=2~452~487.9876(37)+0^{\textnormal{d}}.1366319(6)\times E.
\end{equation}

\begin{figure}
\centering
\rotatebox{270}{\includegraphics[scale=0.4]{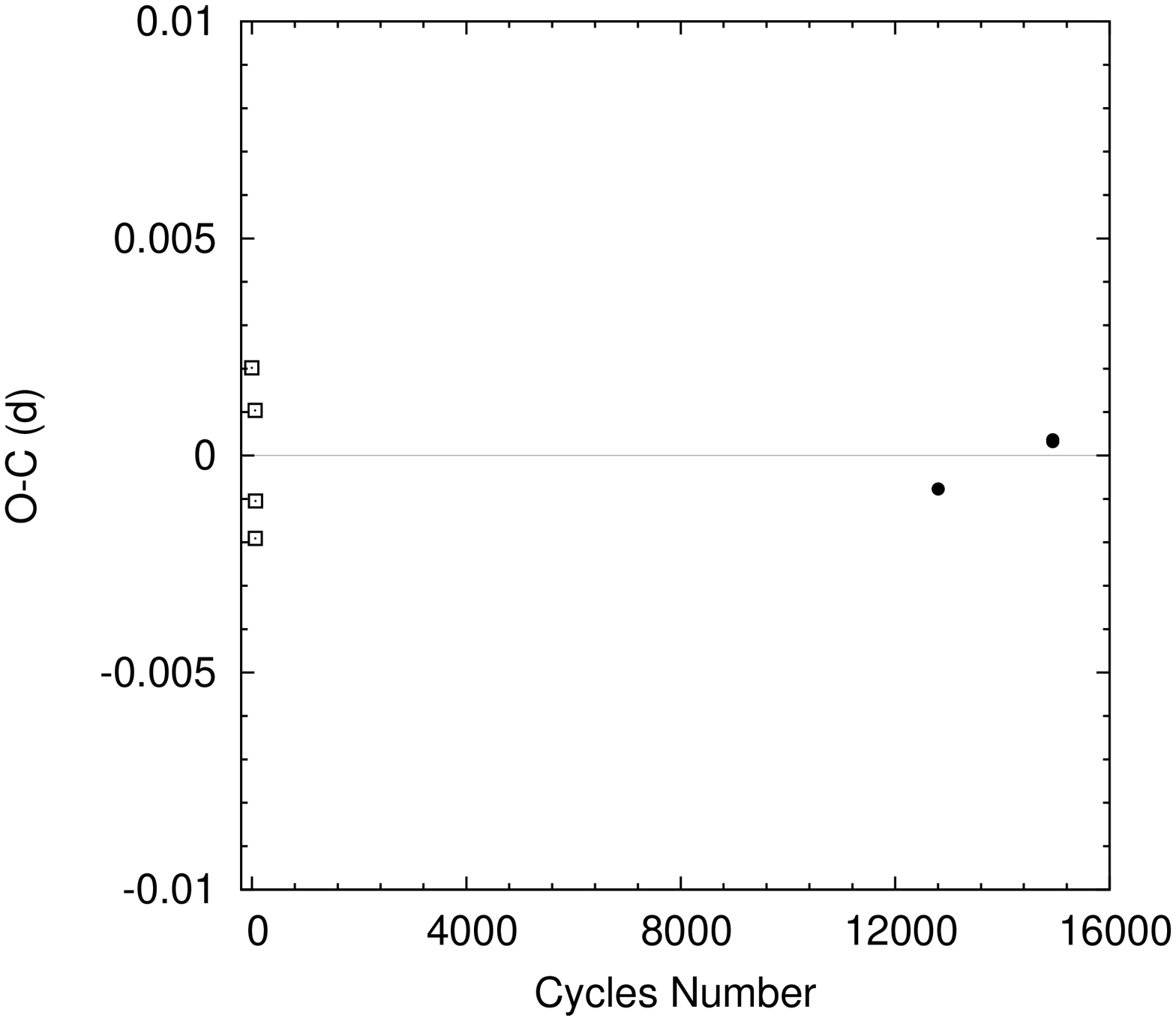}}
\caption{The $O-C$ diagram of SDSS J1327 with respect to the linear ephemeris. The eclipse timings obtained by Wolfe (2003) are shown as open squares, while the new eclipse timings are denoted by solid circles.}
\label{fig3}
\end{figure}

The differences between the observed and calculated times of mid-eclipse are given in Table \ref{tbl-2} and plotted in Fig. \ref{fig3} as a function of the cycle number. The new timings are indicated as solid circles while open squares show eclipse timings taken from Wolfe (2003). It seems that the orbital period of the system appears to be constant during the time span of the observations. The scatter in the \textit{O} -\textit{C} residuals do not exceed a minute. 

\begin{table} 
\footnotesize
\begin{minipage}{8cm}
\caption{Eclipse timings, cycle numbers, \textit{O-C} residuals in unit of day and references for three systems.\label{tbl-2}}
\begin{tabular}{@{}lcr@{.}lcr@{.}lcc@{}}
\hline
Object&HJD (2 400 000+)&\multicolumn{2}{c}{Error}&Cycle Number&\multicolumn{2}{c}{\textit{O-C \textnormal{(d)}}}&Filter&Reference\\
\hline
V849 Oph&48799.7412&&...&0&&0&$V$&Shafter et al. (1993)\\
&48799.9149&&...&1& 0&00095&$V$&"\\
&48831.8736&&...&186&-0&00003&$V$&"\\
&48832.7384&&...&191&0&00100&$V$&"\\
&53962.3846&0&0012&29884&0&03295&$V$&This study  \\
&53963.4197&0&0033&29890&0&03153&$V$&"\\
&53964.2840&0&0006&29895&0&03208&$B$&"\\
J132723.39+652854.2&54237.4101&0&0002&12804&-0&00077&$C$&This study\\
&54529.2551&0&0014&14940&0&00036&$V$&"\\
&54529.3917&0&0003&14941&0&00032&$V$&"\\
&52487.7164&0&002&-2&0&00202&$V$&Wolfe et al. (2003)\\
&52495.7767&0&001&57&0&00104&$V$&"\\
&52496.7301&0&002&64&-0&00191&$B$&"\\
&52496.8676&0&002&65&-0&00105&$B$&"\\
J160745.02+362320.7&53196.7444&&...&-9167& 0&00162&$V$&Szkody et al. (2006)\\
&53219.7335&&...& -9009& 0&00173&$V$&"\\
&54530.5412&0&0001&0&&0&$V$&This study\\
\hline
\end{tabular}
\vspace{-0.5cm}
\end{minipage}
\end{table}
\bigskip 

\subsection{SDSS J1227}
SDSS J1227 is another product of the fifth release of CVs in the SDSS (Szkody et al. 2006). It shows a deep doubling in the higher order Balmer emission lines in the spectra suggesting a high inclination of the system. Also the most prominent spectral features were the broad absorption lines from the white dwarf star. They were taken as indication of low accretion rates that the contribution from an accretion disk is almost negligible. The system is relatively faint, having  a mean magnitude of \textit{g} $\simeq$ 19.10 mag at quiescent. SDSS J1227 was found to be an eclipsing CV with an orbital period of 1.5 h below the period gap. The parameters and evolutionary status of the system were described in details by Littlefair et al. (2008). Furthermore,  amateurs detected the first outburst of the system with an amplitude of ~$\sim$5 mag in 2007 June (actually the exact outburst time is not known). They observed many eclipses during the outburst to decline that present the eclipse profile changes from varying dominated components (Shears et al. 2007). These observations also revealed a super-hump with a period of 0.0653 day, indicating of a SU UMa -type star. There has not been any high-speed photometry during an outburst.

\begin{figure}
\centering
\rotatebox{270}{\includegraphics[scale=0.4]{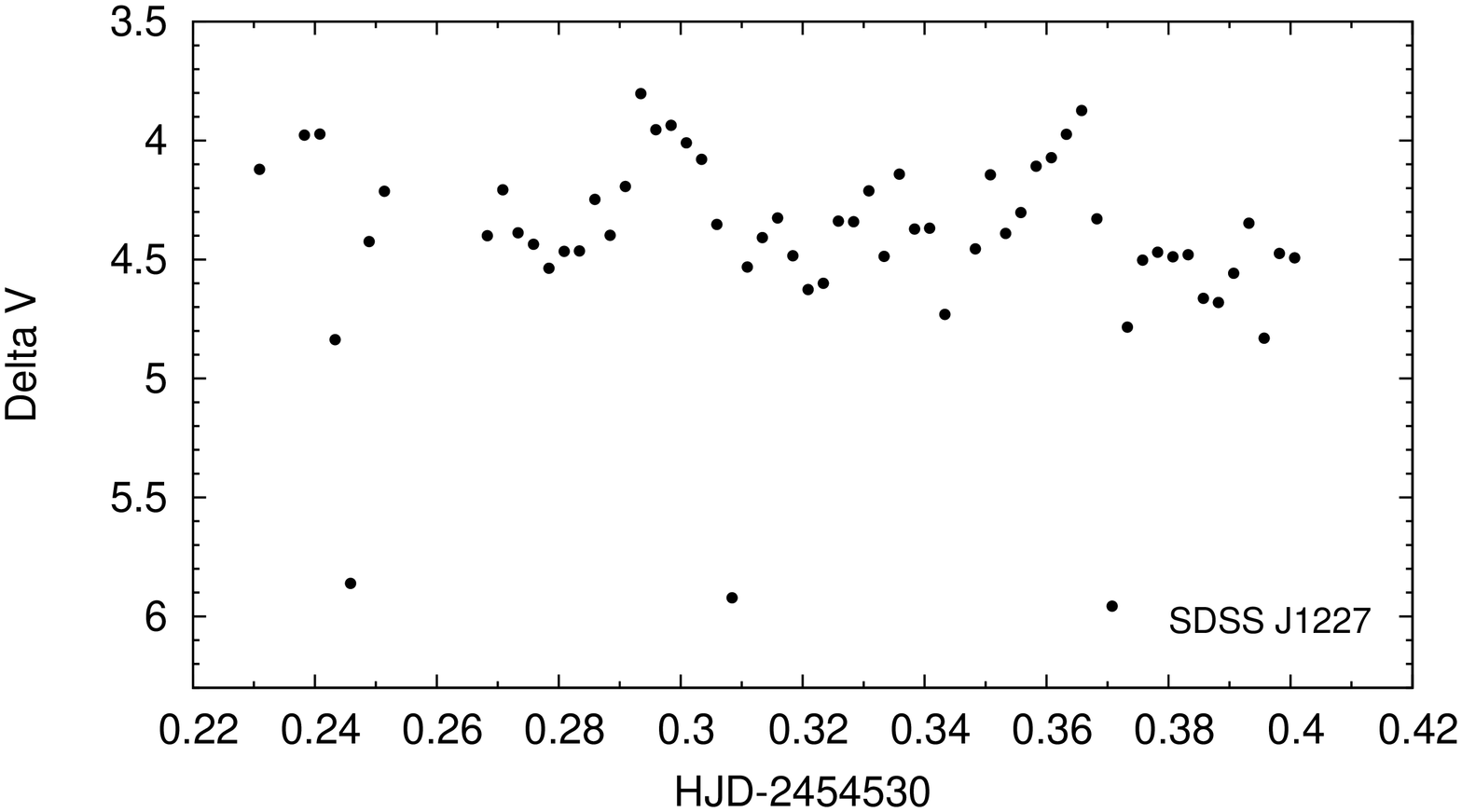}}
\caption{The $V$-band light curve of SDSS J1227 obtained on 2008 March 04. See caption of Fig. \ref{fig1} for a complete description.}
\label{fig4}
\end{figure}

\begin{figure}
\centering
\rotatebox{270}{\includegraphics[scale=0.4]{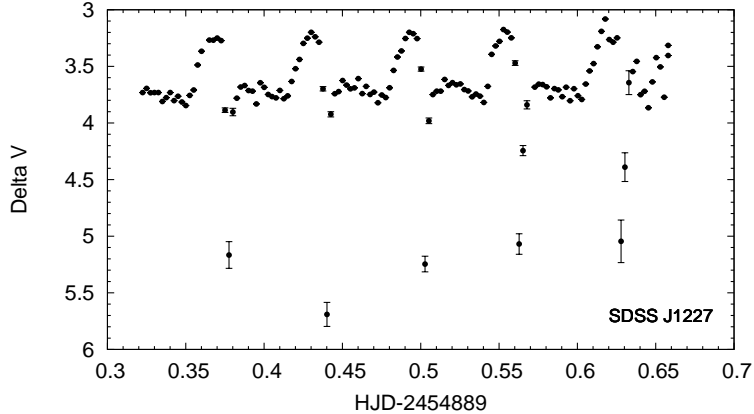}}
\caption{The $V$-band light curve of SDSS J1227  obtained on 2009 February 26. See caption of Fig. \ref{fig1} for a complete description.}
\label{fig5}
\end{figure}

We observed the star photometrically on 2008 March 04 and on 2009 February 26 using the $V$ bandpass. Three eclipses were observed during $\sim$4 h time interval in the first run and the data comprise five eclipses in the second run (Fig. \ref{fig4} and Fig. \ref{fig5}, respectively) . The system has an orbital period of $\sim$90 min and the peak-to-peak light variation is   ~$\sim$2 mag in the  $V$ bandpass. All data were obtained while the system was in quiescence. Since the 200 s integration time was not sufficient to resolve the detailed structure of the eclipse, the actual depth may be slightly larger than we have obtained. The light curve of SDSS J1227 shows prominent orbital humps with an amplitude of $\sim$0.8 mag along with deep eclipses due to a dominant bright spot, lasting approximately half the orbital period alike Z Cha and OY Car. The mid-eclipse timings were measured from the mid-ingress and mid-egress times of the white dwarf eclipse using a combined light curve. Any derivative technique such as described by Wood et al. (1985) could not be used to determine the time of mid-eclipse since we have a very few observations within the eclipse when the star was too faint to measure.  We have therefore employed the following method: to produce a combined light curve with increased time resolution in each cycle, the observation times were shifted in times with respect to a time difference between two observation points in series cycles (i.e. these points are thought to be obtained at the same orbital phase) at the same differential magnitude in the observed light curve. By applying this empirical method, times of the mid-egress and mid-ingress of the white dwarf are become evident.  After that individual eclipse timing was measured from our light curves by using the bisected chord method described in details by Krzeminski (1965) for U Gem. This yields a unique but rough mid-eclipse timing estimation for each cycle. The difference between the measured mid-egress and mid-ingress phases is consistent with the width of the white dwarf eclipse, $\Delta\phi\approx0.040\pm0.005$ in the phase-folded light curve obtained by Littlefair et al. (2008). The mid-eclipse timings have been transformed to the solar system barycentre dynamical time BJDD and listed in Table \ref{tbl-3}. The corresponding uncertainties in the last digit are indicated in the parentheses. Moreover, additional times of minima have been compiled from Littlefair et al. (2008) and Shears et al. (2007) and are given in Table \ref{tbl-3}. 

The \textit{O} -\textit{C} (\textit{I}) are computed using the ephemeris given by Littlefair et al. (2008) as,
\begin{equation}
\textnormal{$Min I$(HJD)}=2~453~796.248245+0^{\textnormal{d}}.062959041\times E.
\end{equation}

\begin{table} 
\footnotesize
\begin{minipage}{8cm}
\caption{Eclipse times of SDSS J1227.\label{tbl-3}}
\begin{tabular}{@{}lcr@{.}lr@{.}lcc@{}}
\hline
BJJD\footnote[1]{Barycentre Dynamical Time }&Cycle Number&\multicolumn{2}{c}{\textit{O-C (I) \textnormal{(d)}}}&\multicolumn{2}{c}{\textit{O-C (II) \textnormal{(d)}}}&Filter\footnote[2]{\textit{$u^{'}$$g^{'}$$r^{'}$}, Sloan Digital Sky Survey colour bands; C, Clear; V, Johnson system }&Reference\\
\hline
2453796.7482445(7)&8&-0&00368&-0&00019&\textit{$u^{'}$$g^{'}$$r^{'}$}&Littlefair et al. (2008)\\
2453797.7554528(8)&24&-0&00381&-0&00016&\textit{$u^{'}$$g^{'}$$r^{'}$}&"\\
2453805.5613010(9)&148&-0&00488&0&00008&\textit{$u^{'}$$g^{'}$$r^{'}$}&"\\
2454264.4689(1)& 7437&-0&00570&-0&00042&$C$&Shears et al. (2007)\\
2454264.5955(1)&7439&-0&00510&0&00026&$C$&"\\
2454264.7219(2)&7441&-0&00460&0&00074&$C$&"\\
2454530.2454(2)&11658&0&02070&-0&00034&$V$&This study\\
2454530.3088(2)&11659&0&02110&0&00011&$V$ &"\\
2454530.3718(2)&11660&0&02110&0&00015&$V$ &"\\
2454889.3770(1)&17363&-0&02910&0&00048&$V$ &"\\
2454889.4396(1)&17364&-0&02940&0&00012&$V$ &"\\
2454889.5022(1)&17365&-0&02980&-0&00023&$V$ &"\\
2454889.5648(1)&17366&-0&03020&-0&00059&$V$ &"\\
\hline
\bigskip
\end{tabular}
\vspace{-0.9cm}
\end{minipage}
\end{table}
\smallskip 
 
\begin{figure}
\centering
\rotatebox{270}{\includegraphics[scale=0.4]{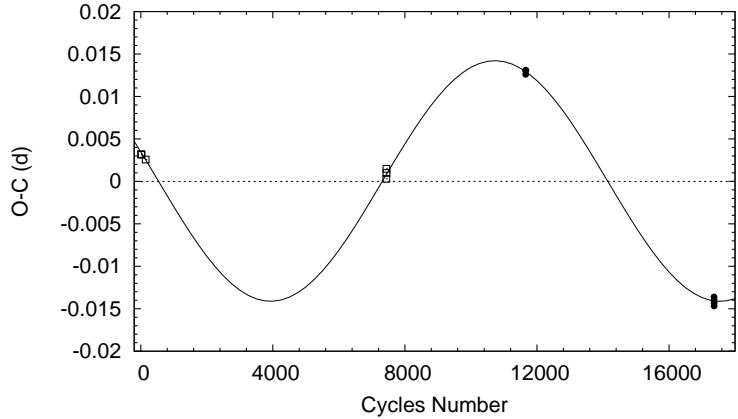}}
\caption{Sinusoidal fit to the $O-C$ residuals of SDSS J1227 is shown as solid line. The eclipse timings from Littefair (2008) and Shears (2007) are shown as open squares, while the new eclipse timings are denoted by solid circles.}
\label{fig6}
\end{figure}

The residuals imply some evidence for a change of the orbital period. The \textit{O} -\textit{C} (\textit{I})  residuals have been plotted against the epoch number in Fig \ref{fig6} which show as if a periodic behaviour. Periodic or sinusoidal-like variations in the  \textit{O} -\textit{C} residuals can be explained by light-time effect resulting from an unseen component in the eclipsing binary or the magnetic activity of the secondary component. If a third body dynamically bounded to SDSS J1227 exists, the eclipsing pair should revolve around the center-of-mass of a three-body system with a period of $2.35\pm0.01$ yr. The projected distance of the center of mass of the eclipsing pair to that of the triple system would be $2.45\pm0.3$ AU. These values lead to a quite large mass function of $f(m_{3})=2.673\pm0.002\:M_{\odot}$ for the hypothetical third body. The mass of such a third body would then range from $4.01\:M_{\odot}$ for $i=90^{\circ}$ to $5.45\:M_{\odot}$ for $i=60^{\circ}$. If the third body having such a large mass were a normal star, it should be bright enough to be detected. Neither spectroscopic observations nor photometric observations of the system show any evidence of the presence of the unseen component star around this binary system.

On the other hand, cyclical orbital period changes are seen in many eclipsing CVs (Borges et al. 2008; Baptista et al. 1995, 2000, 2002, 2003; Wolf et al. 1993; Echevarria\&Alvarez 1993; Bond\&Freeth 1988; Warner 1988). The cycles range from 4 yr in EX Dra to about 30 yr in UX UMa, whereas the amplitude is in the range $10^{1}-10^{2}$ s (Baptista et al. 2003). 
In this case, the variation of the \textit{O} -\textit{C} residuals produced by the magnetic activity  is thought  to be cyclic and can be represented by the following equation: 

\begin{equation}
Min I=T_{0}+P \times E+A_{mod} \sin\left[2 \pi (E-T_{s})/P_{mod}\right]
\end{equation}

\begin{table}
\footnotesize
\begin{minipage}{8cm}
\caption{Result of sinusoidal solution for SDSS J1227 \label{tbl-4}}
\begin{tabular}{@{}lc@{}}
\hline
Parameter&SDSS J1227\\
\hline
$T_{0}(BJD)$&2453796.2414 (2)\\
$P_{orb} (d)$& 0.06295491 (8)\\
$P_{mod} (yr)$&2.35 (1)\\
$asini (AB)$&2.45 (3)\\
\hline
\bigskip
\end{tabular}
\vspace{-0.9cm}
\end{minipage}
\end{table}

where $A_{s}$ is the semi-amplitude, $P_{s}$ the period of sinusoidal variation and $T_{s}$ moment of the minimum of sinusoidal variation. Applying linear least squares fit to the \textit{O} -\textit{C} residuals of SDSS J1227 we determined the unknown parameters in the equation and these parameters are given in Table \ref{tbl-4}. Our result reveals that the orbital period of SDSS J1227 shows cyclical orbital period changes on a time scale of about 2.4 yr with an amplitude $0^{d}.014$. Using these values we find $\Delta P/P=1.04\times10^{-4}$. The deviations from the sinusoidal fit are also given in fourth column of Table \ref{tbl-3} as \textit{O} -\textit{C} (\textit{II}).  However, it should be noted that for a better determination of the cyclical period change the eclipse timings covering at least more than two cycles of the modulation are needed (i.e., at least about a decade of observations). Therefore it is a case of the poor data sampling covering 4 years of observations. However as it is seen from Fig. \ref{fig6} the \textit{O} -\textit{C} residuals exhibit a cyclical period variation. Most probably, the changes of the \textit{O} -\textit{C} (\textit{II}) residuals for SDSS J1227 could be taken as a result of the magnetic activity of the low mass main-sequence component rather than the three-body system.

\subsection{SDSS J1607}
SDSS J1607 was revealed an eclipsing CV by Szkody et al. (2006) since its lines show an indication of doubling structure in higher order Balmer lines. It was found that spectral behaviour of SDSS J1607 resembles a SW Sex type CV (strong CN, He II and a strong single-peaked Balmer H$\alpha$ emission line). They obtained two eclipses in two nights and estimated an orbital period of 3-4 h for the system.

\begin{figure}
\centering
\rotatebox{270}{\includegraphics[scale=0.4]{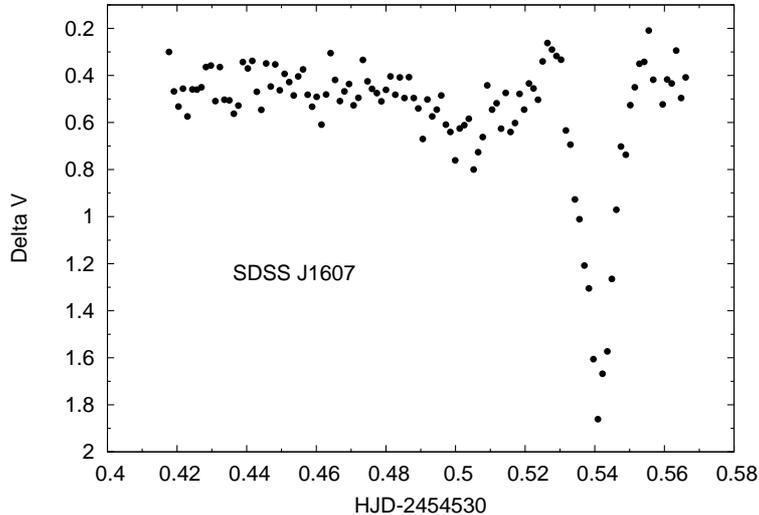}}
\caption{The light curve of SDSS J1607 obtained on 2008 March 4 at quiescence using Johnson $V$ bandpass with 100 sec integration time. It clearly shows  V-shaped eclipse that characteristic feature for a SW Sex star.}
\label{fig7}
\end{figure}

We made photometric observations of the system on 2008 March 4 using Johnson $V$ bandpass with 100 sec integration time. SDSS J1607 showed an eclipse with a depth of 1.5 mag. The eclipse timings  obtained by Szkody et al. (2006) and in this study (Table \ref{tbl-2}) are used to determine orbital period of the system. We derived an orbital period of 3.49 h. The eclipse is fairly symmetrical and there is no evidence of a hot-spot hump preceding the eclipse (Fig. \ref{fig7}). A linear fit to the eclipse timings resulted in the eclipse ephemeris: 

\begin{equation}
\textnormal{$Min I$(HJD)}=2~454~530.5413(1)+0^{\textnormal{d}}.14549982(2)\times E.
\end{equation}

\subsection{SDSS J1457}
\begin{figure}
\centering
\rotatebox{270}{\includegraphics[scale=0.4]{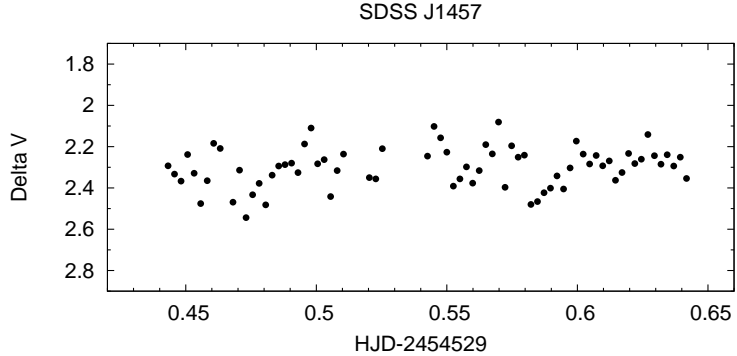}}
\caption{The light curve of SDSS J1457 obtained on 2008 March 3 using the $V$ bandpass.  The light curve does not show a periodic variation that could be attributed to the eclipses but it shows $\sim$40 min quasi-periodic fluctuations with an amplitude  0.4 mag. }
\label{fig8}
\end{figure}

This object is also identified as a CV which candidates for having a high orbital inclination, has a prominent central absorption in Balmer emission line detected by Szkody et al. (2003). SDSS spectra show broad hydrogen absorption lines of a white dwarf which means low accretion rates. SDSS J1457 was included in our observational program due to very little previous information exists. Our time-series observations were made only 4.76 hours on 2008 March 3 in the $V$ bandpass to search for possible eclipses. In spite of $\sim$40 min quasi-periodic variability with an amplitude  0.4 mag, the light curve does not show a periodic variation that could be attributed to the eclipse phenomena (Fig. \ref{fig8}). These variations could be arisen from  flickering at its quiescent phase. However, our observations point out that the orbital period of the system can be much longer than 4.76 h.

\begin{figure}
\centering
\rotatebox{270}{\includegraphics[scale=0.4]{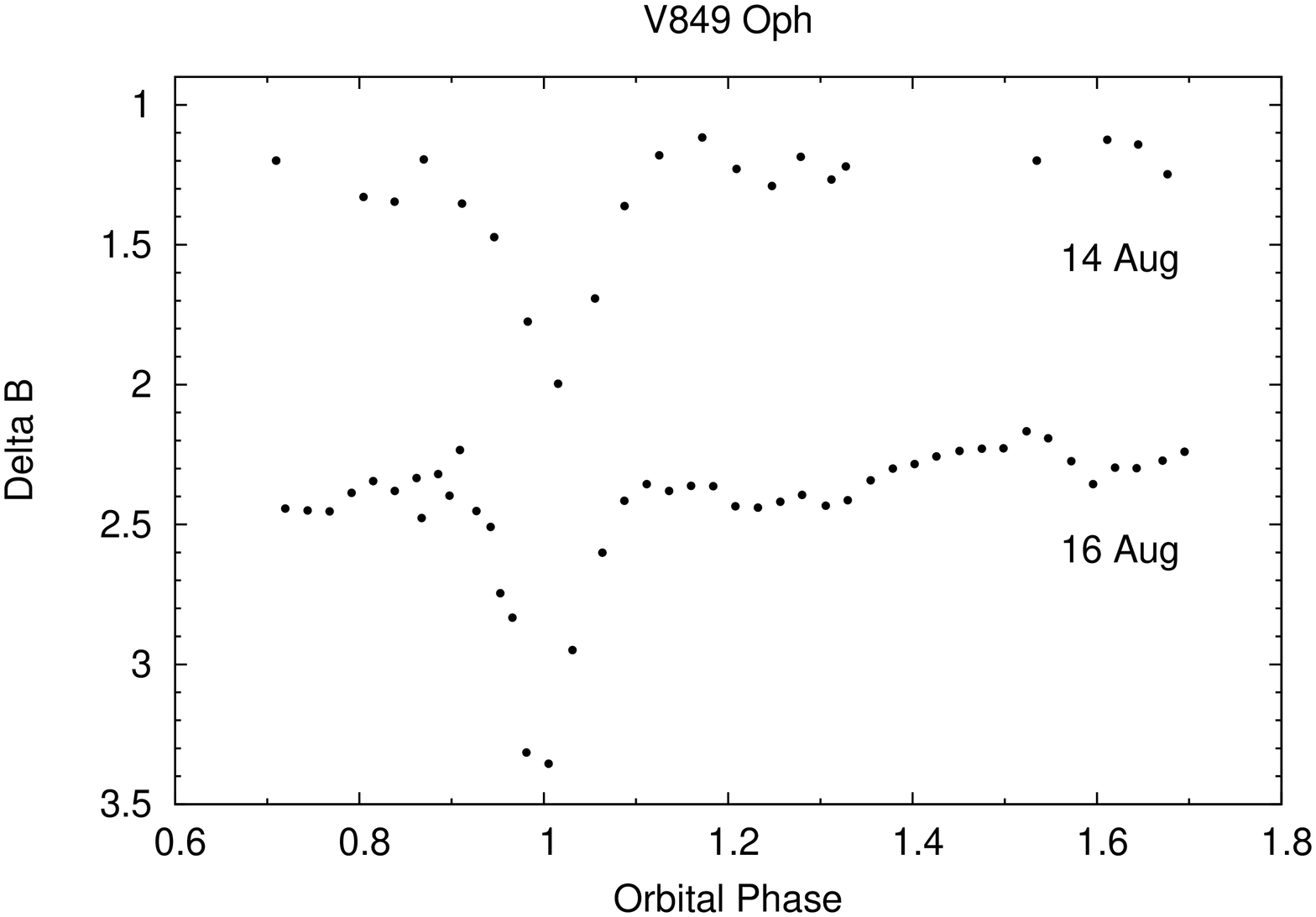}}
\vfill      
\rotatebox{270}{\includegraphics[scale=0.4]{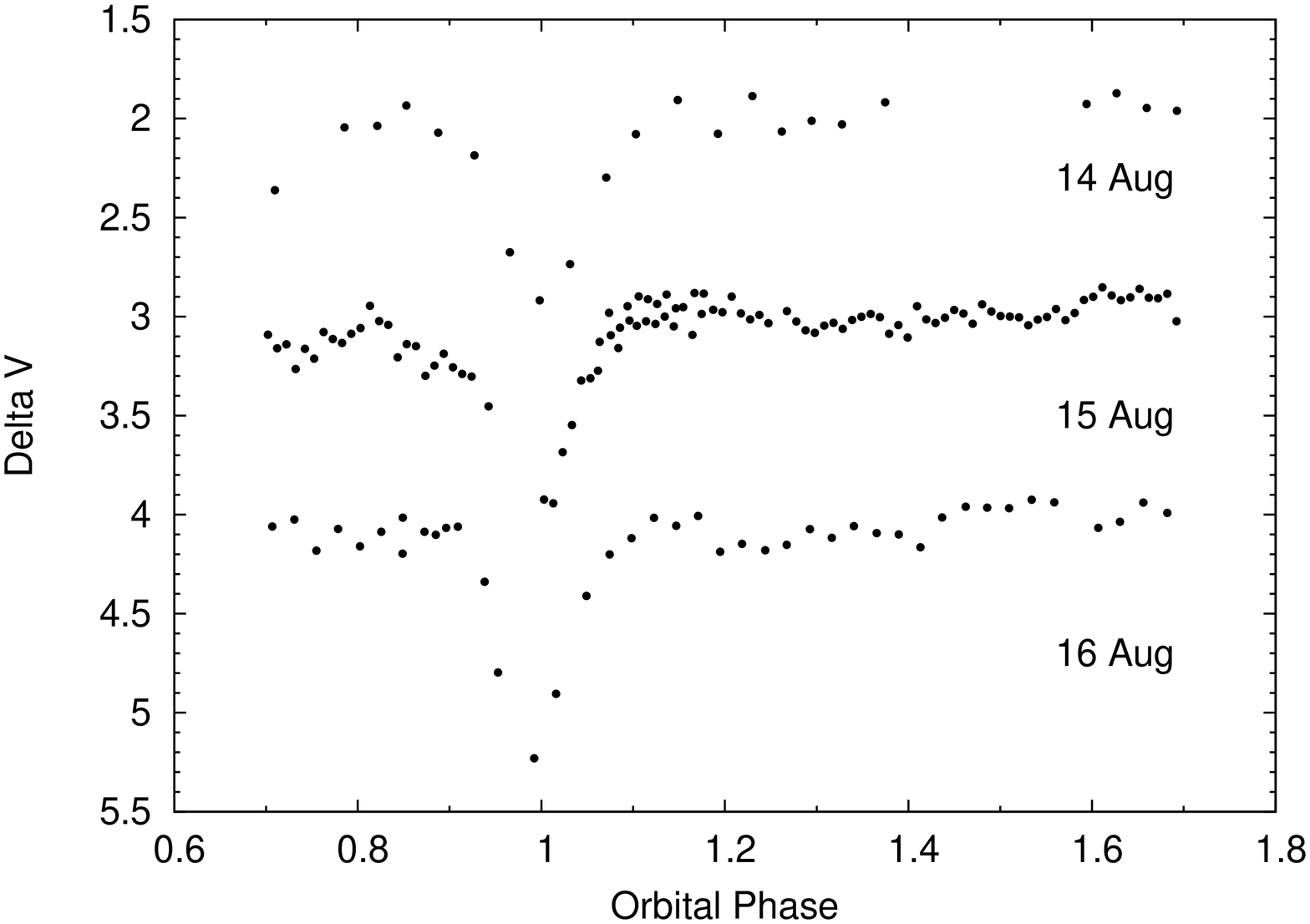}} 
\caption{The light curves of V849 Oph obtained in the $B$ bandpass on 2006 August 14,16 (top) and in the $V$ bandpass (bottom)  on 2006 August 14,15,16 receptively. The upper light curves reflect the real differential magnitudes of V849 Oph. Subsequent light curves are shifted vertically by 1 mag, from top to bottom, for display purposes. The light curve phased according to the ephemeris given in equation (5). }
\label{fig9}
\end{figure}

\subsection{V849 Oph}
V849 Oph was discovered by  Mackie (1919)  and identified as an eclipsing nova (Duerbeck, 1987). Since the brightness of Nova Oph 1919 declined through 3 mag from maximum light in 175 days Duerbeck (1987) classified it as a slow nova. The quiescent visual magnitude of the nova is $\sim$17.9 mag in nowadays. V849 Oph has an eclipse of $\sim$1 mag depth in the $V$ bandpass with an orbital period of 0.172755 days (Shafter et al. 1993). High-speed photometry has not been obtained during an outburst.

Our observations were acquired in Johnson $B$ and $V$ bandpasses on 2006 August 14, 15 and 16. The differential photometric observations obtained using the  $B$ and $V$ bandpasses are plotted versus the orbital phase in the top and bottom panels of  Figure \ref{fig9}. We used 150 s and 120 s integration times, respectively. Eclipses are V-shaped and the eclipse light curve is very smooth without any humps that means the hot spot may not be the main source the radiation. The eclipse timings for V849 Oph are listed in Table \ref{tbl-2}. We refined the orbital period of the system by means of a linear least-squares fit to the eclipse timings given by Shafter et al. 1993 and obtained in this study as follows:

\begin{equation}
\textnormal{$Min I$(HJD)}=2~448~799.7416(3) +0^{\textnormal{d}}.172756(2)\times E
\end{equation}

\section{Conclusion}
We have presented CCD photometry for six faint cataclysmic variable stars at quiescence. Our main goal was to obtain the orbital periods of SDSS CVs using a moderate size telescope and to discuss the preliminary results. The data quality appears  insufficient for revealing the eclipse profile of SDSS J0920, so detailed modelling of the eclipse profile is needed. The change of  $O-C$ residuals for SDSS J1227 with short time coverage of data exhibits a sinusoidal orbital period change that could be arisen as a result of the magnetic activity of the low mass main sequence component. However, new eclipse timings are needed to confirm this variation. Firstly, in this study the orbital period of SDSS J1607 has been found to be above the gap (3.49 h) as suggested by Szkody et al. (2006). The high resolution time-resolved spectra will identify whether it is a SW Sex type of CV and confirm if there is an offset between photometric and spectroscopic phases (typically $\sim$0.2). SDSS J1327 shows symmetric, large amplitude (2.5 mag) eclipses with an orbital period of 3.28 h. In addition, as a result of our observation we suggest that either  the brightness of SDSS J1457 is constant or the orbital period is much longer than 4.76 h. Further high speed differential multi-colour photometry of these CVs with large-telescopes are needed in order to reveal the true nature of the light variations.

\section*{Acknowledgements} We thank the anonymous referees for the helpful comments and suggestions on an earlier draft of this manuscript. We also would like to thank TUG (T\"{U}B\.{I}TAK National Observatory) and T\"{U}B\.{I}TAK-BAYG for their supports. This study has been partly supported by Ege University Research Project (2006/FEN/003).

\clearpage


\begin{thebibliography}{99}

\bibitem[]{} Baptista R., Horne K., Hilditch R., Mason K.O., Drew J.E., 1995, ApJ, 448, 395
\bibitem[]{} Baptista R., Catalan M.S.Costa L., 2000, MNRAS, 316, 529
\bibitem[]{} Baptista R., Jablonski F.J., Oliveira E., et al., 2002, MNRAS, 335, L75
\bibitem[]{} Baptista R., Borges B.W., Bond H.E., et al., 2003, MNRAS, 345, 889
\bibitem[]{} Bond I.A., Freeth R.V., 1988, MNRAS, 232, 753
\bibitem[]{} Borges B.W., Baptista R., Papadimitriou C., Giannakis O., 2008, A\&A, 480, 481
\bibitem[]{}Dillon M., G\"{a}nsicke B.T., Aungwerojwit A., Rodriguez-Gill P., Marsh T.R. et al. ,2008, MNRAS, 386, 1568
\bibitem[]{}Duerbeck H. W. , 1987, ApJ, Space. Sci. Rev, 437, 879
\bibitem[]{} Echevarria J.,  Alvarez M., 1993, , 214, A\&A, 275, 187
\bibitem[]{}G\"{a}nsicke B.T., Hagen H.J., \& Engels  D., 2002, ASP Conf. Ser. Vol. 261,190 
\bibitem[]{}Groot  P. J., Vreeswijk  P. M.; Huber  M. E. et al. 2003, MNRAS, 339, 427-434
\bibitem[]{}Howell S.B., Szkody P., 1988, PASP, 100, 224
\bibitem[]{}Howell S.B., Szkody P., Mateo M., Kreidl T. 1989, PASP, 101, 899
\bibitem[]{}Howell S.B., Szkody P., Kreidl T., Mason K., Puchnarewichz E.M.,1990, PASP, 102, 758
\bibitem[]{}Howell S.B., Szkody P., Kreidl T., Dobrzycka D.,1991, PASP, 103, 300
\bibitem[]{} Krzeminski  W., 1965, ApJ, 142, 1051
\bibitem[]{}Littlefair  S. P.,Dhillon  V. S., Marsh  T. R., G\"{a}nsicke  B. T., Southworth J., Baraffe  I., Watson  C. A., Copperwheat  C., 2008, MNRAS, 388, 1582-1594
\bibitem[]{}Schmidth G.D., Szkody P., Henden  A, Anderson  S. F., Lamb D. Q., Margon B., Schneider D.P., 2007, ApJ, 654, 521
\bibitem[]{}Shafter A.W., Misselt K.A., Veal J.M., 1993, PASP, 105, 853
\bibitem[]{}Shears J., Brandy S., Foote J., Starkey D., Vanmunster T., 2007, e-print arXiv, 0711, 2136
\bibitem[]{}Szkody P., Anderson S. F., Ag\"{u}eros M., Covarrubias R., Bentz M., Hawley S., Margon B. et al., 2002, AJ, 123, 430
\bibitem[]{}Szkody P., Fraser O., Silvestri N., Henden  A., Anderson, S. F., Frith J., Lawton B. et al., 2003, AJ, 126, 1499
\bibitem[]{}Szkody P., Henden  A., Fraser O., Silvestri N., Bochanski J. J., et al., 2004, AJ, 128,1882
\bibitem[]{}Szkody P., Henden, A., Fraser O., Silvestri N., Schmidt G.D., Bochanski J. J., Wolfe M.A., et al., 2005, AJ, 129, 2386
\bibitem[]{}Szkody P., Henden  A., Ag\"{u}eros M., Anderson, S. F., Bochanski J. J., Knapp G. R., Mannikko L., et al. 2006, AJ, 131, 973
\bibitem[]{}Szkody P., Henden  A., Mannikko L., Mukadam A., Schmidt G.D., Bochanski J. J., Ag\"{u}eros M., et al. 2007, AJ, 134, 185
\bibitem[]{}Warner  B., 1988, Nat, 336, 129
\bibitem[]{}Wolf S., Mantel K.H., Horne K., et al., 1993, A\&A, 275, 187
\bibitem[]{}Wolfe  M. A.; Szkody, P.; Fraser  O. J.; Homer  L.; Skinner  S.; Silvestri  N. M., 2003, PASP, 115, 811, 1118-1123
\bibitem[]{}Woudt P.A., Warner B., 2001, MNRAS, 328, 159
\bibitem[]{}Woudt P.A., Warner B., 2002, MNRAS, 335, 44
\bibitem[]{}Woudt P.A., Warner B., 2003, MNRAS, 340, 1011
\bibitem[]{}Woudt P.A., Warner B., Pretorius M.L., 2004, MNRAS, 351, 1015
\bibitem[]{}Woudt P.A., Warner B., Spark M., 2005, MNRAS, 364, 107
\bibitem[]{} Wood J.H., Irwin M.J., Pringle J.E., 1985, MNRAS, 214, 475
\end{thebibliography}
\end{document}